# Structure-Behavior Coalescence Process Algebra

## --Toward a Unified View of the System in Model-Based Systems Engineering

William S. Chao

*Abstract*—In **Model-Based Systems Engineering (MBSE), the Systems Modeling Language (SysML) specification includes a metamodel that defines the language concepts and a user model that defines how the language concepts are represented. In SysML, an important use of metamodel is to provide an integrated semantic framework that every diagram in the user model can be projected as a view of the metamodel. However, most existing SysML metamodels lack such capability of being a basis for unification of different views of a system. To overcome the shortcomings of the current SysML metamodel approaches, we developed Channel-Based Multi-Queue Structure-Behavior Coalescence Process Algebra (C-M-SBC-PA), which provides an integrated semantic framework that is able to integrate structural constructs with behavioral constructs. Using C-M-SBC-PA as the metamodel of SysML, each diagram in the user model can be projected as a view of the C-M-SBC-PA metamodel.**

## I. INTRODUCTION

As a systems modeling language for Model-Based Systems Engineering (MBSE) applications [1], the Systems Modeling Language (SysML) [2]-[3] defines a set of language concepts that are used to model the structure and behavior of a system. The SysML concepts include (a) an abstract syntax that defines the language concepts and is described by a metamodel, and (b) a concrete syntax, or notation, that defines how the language concepts are represented and is described by a user model [4]-[5].

Since SysML is a multi-diagram approach, there are always some inconsistencies between different diagrams in the user model [6]-[9]. To ensure and check the consistency, a metamodel that defines the abstract syntax of a modeling language needs to provide an integrated semantic framework for defining consistency rules to impose constraints on the structure or behavior constructs. It is hoped that through this integrated semantic framework, each diagram in the user model can be projected as a view of the metamodel.

Unfortunately, most current SysML metamodels are not able to project every diagram in the user model as a view of the metamodel. In this paper, we developed Channel-Based Multi-Queue Structure-Behavior Coalescence Process Algebra (C-M-SBC-PA) [10]-[11] as a metamodel of SysML. In C-M-SBC-PA, each diagram in the user model will be projected as a view of the metamodel. Therefore, we claim that C-M-SBC-PA genuinely provides a basis for unification of different views of the system in Model-Based Systems Engineering.

The remainder of this paper is arranged as follows. Section 2 deals with the current SysML metamodel study. C-M-SBC-PA as a metamodel for SysML is detailed in Section 3. After detailing C-M-SBC-PA, we will validate the SBC method with a case study in Section 4. Conclusions of this paper are in Section 5.

## II. RELATED STUDIES

. In SysML, a metamodel is used to describe the concepts in the language, their characteristics, and interrelationships. This is sometimes called the abstract syntax of the language, and is distinct from the concrete syntax that specifies the user model for the language. A significant usage of the metamodel is to provide an integrated semantic framework that every diagram in the user model can be projected as a view of the metamodel.

The Object Management Group (OMG) defines a language for representing metamodes, called Meta Object Facility (MOF) that is used to define UML, SysML and other metamodels. Various mechanisms are used in MOF, such as Object Constraint Language (OCL) [12], Foundational UML (fUML) [13], The Action Language for Foundational UML (Alf) [14], Process Specification Language (PSL) [15], to name a few.

The Object Constraint Language is a precise text language that provides constraint on the structure (i.e., blocks) to ensure consistency of the user model [12]. However, not every diagram in the user model can be projected as a view of the OCL metamodel because the OCL fails to provide a unified core semantics framework. Therefore, the OCL metamodel can only ensure part of user model consistency.

The Foundational UML is a subset of the standard UML for which a standard execution constraint language, PSL, is used to define the semantics of the execution model [13]. Although fUML provides constraint on the behavior (i.e., activities) to make the model executable, it fails to integrate the structural constructs with the behavioral constructs. Not being able to provide a unified core semantics framework, the Foundational UML can not project every diagram in the user model as a view of the fUML metamodel.

The Action Language for Foundational UML (Alf) is a complementary specification to Foundational UML [14]. The key use of Alf is to act as the notation for specifying executable behaviors in SysML, for example, methods for object operations, the behavior of a block, or transition effects on state machines. Like fUML, Alf also fails to provide a unified core semantics framework to integrate the structural constructs with the behavioral constructs. Therefore, the Alf is not able to project every diagram in the user model as a view of the Alf metamodel.

In order to overcome the shortcomings of the current SysML metamodel approaches, we need to develop an integrated semantics framework that is able to unify structural constructs with behavioral constructs. Channel-Based Multi-Queue Structure-Behavior Coalescence Process Algebra (C-M-SBC-PA) is such a candidate. In C-M-SBC-PA, the structural and behavioral constructs are unified. Using C-M-SBC-PA as the integrated semantics framework for SysML, each diagram in the user model will be projected as a view of the C-M-SBC-PA metamodel.

In order to overcome the shortcomings of the current SysML metamodel approaches, we need to develop an integrated semantics framework that is able to unify structural constructs with behavioral constructs. Channel-Based Multi-Queue Structure-Behavior Coalescence Process Algebra (C-M-SBC-PA) is such a candidate. In C-M-SBC-PA, the structural and behavioral constructs are unified. Using C-M-SBC-PA as the integrated semantics framework for SysML, each diagram in the user model will be projected as a view of the C-M-SBC-PA metamodel.

## III. METHOD OF CHANNEL-BASED MULTI-QUEUE SBC PROCESS ALGEBRA

### A. Channel-Based Value-Passing Interactions

The block is the fundamental modular unit for describing systems structure in SysML [4]-[5]. It can define a type of conceptual or physical entity; a hardware, software, or data component; a person; a facility; or an entity in the natural world. A channel is a mechanism for agent communication via message passing [10]-[11]. A message may be sent over a channel, and another agent is able to receive messages sent over a channel it has a reference to. Each channel defines a set of parameters that describes the arguments passed in with the request, or passed back out once a request has been handled. The signature for a channel is a combination of its name along with parameters as follows:

**<channel name> ( )**

The parameters in the parameter list represent the inputs or outputs of the channel. Each parameter in the list is displayed with the following format:

**<direction> : **

Parameter direction may be in, out, or inout. We formally describe the "channel signature" as a relation $K \subseteq \Lambda \times \Theta$ where $\Lambda$ is a set of "channel names" and $\Theta$ is a set of "parameter lists".

An interaction [11]-[12] represents an indivisible and instantaneous communication or handshake between the caller agent (either external environment's actor or block) and the callee agent (block). In the channel-based value-passing approach as shown in Figure 1, the caller agent interacts with the callee block through the channel interaction. In the figure, getPastDueBalance(in studentId: String; out PastDueBalance: Real) is a channel signature. Figure 1 also depicts that the "getPastDueBalance" channel is **required** by the caller and is **provided** by the callee block.

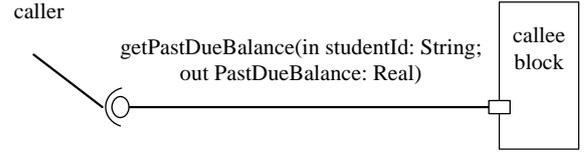

Figure 1.  Channel-based Value-passing Interaction

The external environment uses a "type 1 interaction" to interact with a block. We formally describe the channel-based value-passing "type 1 interaction" as a relation $G \subseteq B \times K \times \Gamma$ where $B$ is a set of "external environment's actors" and $K$ is a set of "channel signatures" and $\Gamma$ is a set of "blocks".

Two blocks use a "type 2 interaction" to interact with each other. We formally describe the channel-based value-passing "type 2 interaction" as a relation $V \subseteq \Gamma \times K \times \Gamma$ where $\Gamma$ is a set of "blocks" and $K$ is a set of "channel signatures".

We can also formally describe the channel-based value-passing "type 1 or 2 interaction" as a relation $\Delta \subseteq \Xi \times K \times \Gamma$ where $\Xi$ is a set of "external environment's actors or blocks" and $K$ is a set of "channel signatures" and $\Gamma$ is a set of "blocks".

### B. Entities of C-M-SBC-PA

As shown in Table I, we assume a relation $K$ of channel signatures, and use $k_1, k_2...$ to range over $K$. Further, we let $\Lambda$ be the set of channel names, and use $ch_1, ch_2...$ to range over $\Lambda$. We let $\Theta$ be the set of parameters, and use $p_1, p_2...$ to range over $\Theta$. We let $B$ be the set of actors, and use $\beta_1, \beta_2...$ to range over $B$. We let $\Gamma$ be the set of blocks, and use $b_1, b_2...$ to range over $\Gamma$. We assume $\Xi$ be the set of actors or blocks, and use $\rho_1, \rho_2...$ to range over $\Xi$. We let $G$ be the relation of type 1 interactions, and use $g_1, g_2...$ to range over $G$. We let $V$ be the relation of type 2 interactions, and use $v_1, v_2...$ to range over $V$. Further, we let $\Delta$ be a set of type 1 or 2 interactions, and employ $a_1, a_2...$ to range over $\Delta$. We assume $\Psi$ be the set of state expressions, and employ $s_1, s_2...$ to range over $\Psi$. Further, we assume $\Phi$ be the set of state constants, and employ $A_1, A_2...$ to range over $\Phi$.

TABLE I
ENTITIES OF C-M-SBC-PA

| Entity Set | Entity Name | Entity Type |
|---|---|---|
| $K$ | $k_1, k_2...$ | channel signatures |
| $\Lambda$ | $ch_1, ch_2...$ | channel names |
| $\Theta$ | $p_1, p_2...$ | parameter lists |
| $B$ | $\beta_1, \beta_2...$ | actors |
| $\Gamma$ | $b_1, b_2...$ | blocks |
| $\Xi$ | $\varrho_1, \varrho_2...$ | actors or blocks |
| $G$ | $g_1, g_2...$ | type 1 interactions |
| $V$ | $v_1, v_2...$ | type 2 interactions |
| $\Delta$ | $a_1, a_2...$ | type 1 or 2 interactions |
| $\Psi$ | $s_1, s_2...$ | states |
| $\Phi$ | $A_1, A_2...$ | state constants |

### C. SBC Interaction Transition Graph

In Channel-Based Multi-Queue Structure-Behavior Coalescence Process Algebra, we use the SBC interaction transition graph (ITG) as a single diagram to specify the semantics of a system. The SBC interaction transition graph is a labelled transition system (LTS) [16]. Overall, the SBC interaction transition graph provides the integrated semantics framework for the C-M-SBC-PA method to integrate structural constructs with behavioral constructs. In C-M-SBC-PA, each state is regarded as a process. The notion of a SBC interaction transition graph is defined as follows.

**DEFINITION** (INTERACTION TRANSITION GRAPH) A SBC interaction transition graph $ITG = (\Psi, s_0, \Xi, \Lambda, \Theta, \Gamma, ITGR)$ consists of

- *a finite set $\Psi$ of states,*

- *an initial state $s_0 \in \Psi$,*

- *a finite set $\Xi$ of external environment's actors or blocks,*

- *a finite set $\Lambda$ of channel names,*

- *a finite set $\Theta$ of parameter lists,*

- *a finite set $\Gamma$ of blocks,*

- *a transition relation $ITGR \subseteq \Psi_1 \times \Xi \times \Lambda \times \Theta \times \Gamma \times \Psi_2$,*

  where $(s_j, \varrho, ch, p, b, s_k) \in ITGR$ *is written*

as $s_j \xrightarrow{\varrho, ch, p, b} s_k$ .

We can draw a diagram to represent the SBC interaction transition graph. Figure 2 shows the diagram of the SBC interaction transition graph $ITG_{01}$. In the diagrammed SBC interaction transition graph, the state is represented by a circle containing its name; the transition from the source state to the target state is represented by an arrow and labelled with an interaction; the initial state (for example, "$s_1$") is the target state of the transition that has no source state. In a state, if multiple transitions to be triggered are met, the choice of trigger will be arbitrary and fair.

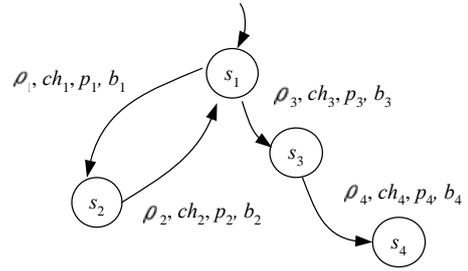

Figure 2. Diagram of the Interaction Transition Graph $ITG_{01}$

We can also list the relationships that represent the SBC interaction transition graph. Table II shows the transition relation $ITGR_{01}$ of the SBC interaction transition graph $ITG_{01}$.

TABLE II
Relation $ITGR_{01}$ of the Transition Graph $ITG_{01}$

| $\Psi_1$ | $\Xi$ | $\Lambda$ | $\Theta$ | $\Gamma$ | $\Psi_2$ |
|---|---|---|---|---|---|
| $s_1$ | $\varrho_1$ | $ch_1$ | $p_1$ | $b_1$ | $s_2$ |
| $s_2$ | $\varrho_2$ | $ch_2$ | $p_2$ | $b_2$ | $s_1$ |
| $s_1$ | $\varrho_3$ | $ch_3$ | $p_3$ | $b_3$ | $s_3$ |
| $s_3$ | $\varrho_4$ | $ch_4$ | $p_4$ | $b_4$ | $s_4$ |

In order to reduce the complexity of the SBC interaction transition graph, we shall introduce an orthogonal composite state. An orthogonal composite state in the SBC interaction transition graph may have many regions, which may each contain substates. These regions are orthogonal to each other. When an orthogonal composite state is active, each region has its own active state that is independent of the others and any incoming interaction is independently analyzed within each region. We use $ITG_1 \| ITG_2 \| ITG \| ... \| ITG_m$ to represent an orthogonal composite state, which means the composition of $ITG_1, ITG_2, ITG_3, ...,$ and $ITG_m$.

In C-M-SBC-PA, the SBC interaction transition graph of the system $ITG_{system}$ is defined as $\|_{i=1, m} ITG_i$ or $ITG_1 \parallel ITG_2 \parallel \dots \parallel ITG_m$. Each SBC interaction transition graph $ITG_i$ is represented by the transition relation $ITGR_i \subseteq \Psi_1$ X $\Xi$ X $\Lambda$ X $\Theta$ X $\Gamma$ X $\Psi_2$, where $(s_{ij}, \rho, ch, p, b, s_{ik}) \in ITGR_i$ is denoted by $s_{ij} \xrightarrow{\rho, ch, p, b} s_{ik}$. The SBC interaction transition graph of the system $ITG_{system}$ is represented by the transition relation $ITGR_{system}$ which is defined as $\parallel_{i=1, m} ITGR_i$ or $ITGR_1 \parallel ITGR_2 \parallel \dots \parallel ITGR_m$.

We can draw a diagram to represent the SBC interaction transition graph of a system. Figure 3 shows a diagram of the SBC interaction transition graph $ITG_{system}$.

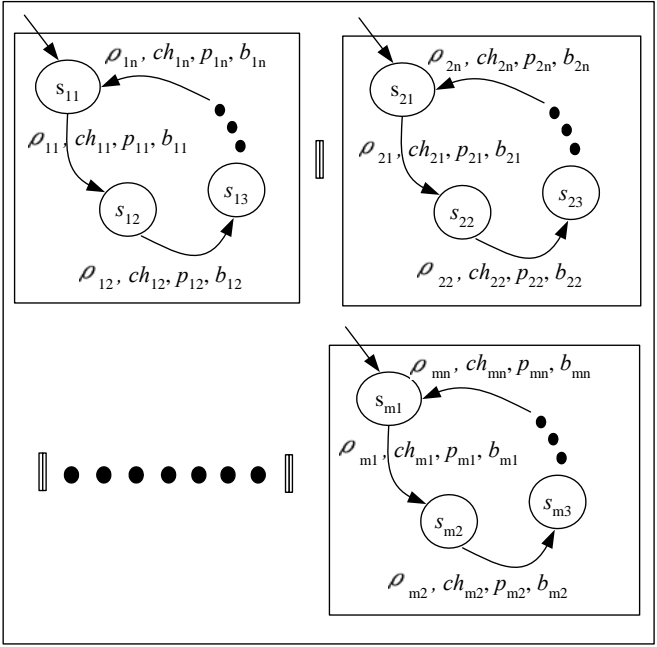

Figure 3. Transition Graph $ITG_{system}$

We can also list the relationships that represent the SBC interaction transition graph of a system. Table III shows the transition relation $ITGR_{system}$ of the SBC interaction transition graph $ITG_{system}$.

| $\Psi_1$ | $\Xi$ | $\Lambda$ | $\Theta$ | $\Gamma$ | $\Psi_2$ |
|---|---|---|---|---|---|
| $s_{11}$ | $\rho_{11}$ | $ch_{11}$ | $p_{11}$ | $b_{11}$ | $s_{12}$ |
| $s_{12}$ | $\rho_{12}$ | $ch_{12}$ | $p_{12}$ | $b_{12}$ | $s_{13}$ |
| • | • | • | • | • | • |
| $s_{1n}$ | $\rho_{1n}$ | $ch_{1n}$ | $p_{1n}$ | $b_{1n}$ | $s_{11}$ |

| $\Psi_1$ | $\Xi$ | $\Lambda$ | $\Theta$ | $\Gamma$ | $\Psi_2$ |
|---|---|---|---|---|---|
| $s_{21}$ | $\rho_{21}$ | $ch_{21}$ | $p_{21}$ | $b_{21}$ | $s_{22}$ |
| $s_{22}$ | $\rho_{22}$ | $ch_{22}$ | $p_{22}$ | $b_{22}$ | $s_{23}$ |
| • | • | • | • | • | • |
| $s_{2n}$ | $\rho_{2n}$ | $ch_{2n}$ | $p_{2n}$ | $b_{2n}$ | $s_{21}$ |

| $\Psi_1$ | $\Xi$ | $\Lambda$ | $\Theta$ | $\Gamma$ | $\Psi_2$ |
|---|---|---|---|---|---|
| $s_{m1}$ | $\rho_{m1}$ | $ch_{m1}$ | $p_{m1}$ | $b_{m1}$ | $s_{m2}$ |
| $s_{m2}$ | $\rho_{m2}$ | $ch_{m2}$ | $p_{m2}$ | $b_{m2}$ | $s_{m3}$ |
| • | • | • | • | • | • |
| $s_{mn}$ | $\rho_{mn}$ | $ch_{mn}$ | $p_{mn}$ | $b_{mn}$ | $s_{m1}$ |

### D. Projecting the SysML Internal Block Diagram from the SBC Interaction Transition Graph

In model-based systems engineering, the internal block diagram shows the connection between parts of a block. The notion of an internal block diagram is defined as follows.

**DEFINITION** (INTERNAL BLOCK DIAGRAM) A SysML internal block diagram $IBD = (\Xi, \Lambda, \Theta, \Gamma, IBDR)$ consists of

- *a finite set $\Xi$ of external environment's actors or blocks*,

- *a finite set $\Lambda$ of channel names*,

- *a finite set $\Theta$ of parameter lists*,

- *a finite set $\Gamma$ of blocks*,

- *a relation $IBDR \subseteq \Xi$ X $\Lambda$ X $\Theta$ X $\Gamma$, where $(\rho, op, p, b) \in IBDR$*.

The SysML internal block diagram of the system $IBD_{system}$ is represented by the relation $IBDR_{system} \subseteq \Xi$ X $\Lambda$ X $\Theta$ X $\Gamma$, where $(\rho, op, p, b) \in IBDR$, as shown in Table IV.

| $\Xi$ | $\Lambda$ | $\Theta$ | $\Gamma$ |
|---|---|---|---|
| $\rho_1$ | $ch_1$ | $p_1$ | $b_1$ |
| $\rho_2$ | $ch_2$ | $p_2$ | $b_2$ |
| $\rho_3$ | $ch_3$ | $p_3$ | $b_3$ |
| $\rho_4$ | $ch_4$ | $p_4$ | $b_4$ |
| • | • | • | • |
| $\rho_m$ | $ch_m$ | $p_m$ | $b_m$ |

The algorithm used to project the IBD relation $IBDR_{system} \subseteq \Xi \times \Lambda \times \Theta \times \Gamma$ from the ITG relation $ITGR_{system} = \Psi_1 \times \Xi \times \Lambda \times \Theta \times \Gamma \times \Psi_2$ is as follows.

**ALGORITHM 1** (Projecting $IBDR_{system}$ from $ITGR_{system}$)

  **For** i = 1, m **Loop**

   SELECT $\Xi$, $\Lambda$, $\Theta$, $\Gamma$ INTO $IBDR_i$ ($\Xi$, $\Lambda$, $\Theta$, $\Gamma$) FROM $ITGR_i$ ;

   INSERT INTO $IBDR_{1~m}$ ($\Xi$, $\Lambda$, $\Theta$, $\Gamma$) SELECT * FROM $IBDR_i$

  **End Loop;**

  SELECT DISTINCT * INTO $IBDR_{system}$ FROM $IBDR_{1~m}$

**END ALGORITHM**

Once we have the IBD relation $IBDR_{system}$, it is easy to get a SyML internal block diagram of the system.

### E. Projecting the SysML State Machine Diagram from the SBC Interaction Transition Graph

In model-based systems engineering, the state machine diagram represents behavior of a system in terms of its transition between states triggered by actions. The notion of the state machine diagram is defined as follows.

**DEFINITION** (STATE MACHINE DIAGRAM) A SysML state machine diagram $SMD = (\Psi, \Lambda, SMDR)$ consists of

- *a finite set $\Psi$ of states,*

- *a finite set $\Lambda$ of channel names,*

- *a relation $SMDR \subseteq \Psi_1 \times \Lambda \times \Psi_2$,*
  where $(s_j, ch, s_k) \in SMDR$ *is denoted by* $s_j \xrightarrow{ch} s_k$ .

The SysML state machine diagram of the system $SMD_{system}$ is defined as $\Vert_{i=1,m} SMD_i$ or $SMD_1 \Vert SMD_2 \Vert \ldots \Vert SMD_m$. Each state machine diagram $SMD_i$ is represented by the relation $SMDR_i \subseteq \Psi_1 \times \Lambda \times \Psi_2$, where $(s_{ij}, ch, s_{ik}) \in SMDR_i$ is denoted by $s_{ij} \xrightarrow{ch} s_{ik}$ . The state machine diagram of the system $SMD_{system}$ is represented by the relation $SMDR_{system}$ which is defined as $\Vert_{i=1,m} SMDR_i$ or $SMDR_1 \Vert SMDR_2 \Vert \ldots \Vert SMDR_m$, as shown in Table V.

TABLE V
Relation $SMDR_{system}$

The algorithm used to project the SMD relation $SMDR_{system} \subseteq \Psi_1 \times \Lambda \times \Psi_2$ from the ITG relation $ITGR_{system} = \Psi_1 \times \Xi \times \Lambda \times \Theta \times \Gamma \times \Psi_2$ is as follows.

**ALGORITHM 2** (Projecting $SMDR_{system}$ from $ITGR_{system}$)

  **For** i = 1, m **Loop**

   SELECT $\Psi_1$, $\Lambda$, $\Psi_2$ INTO $SMDR_i$ FROM $ITGR_i$ ;

  **End Loop;**

  ORTHOGONALLY COMPOSE ALL $SMDR_i$ (i.e., $\Vert_{i=1,m} SMDR_i$ ) TO GET $SMDR_{system}$

**END ALGORITHM**

Once we have the SMD relation $SMDR_{system}$, it is easy to get the SyML state machine diagram of the system.

### F. Projecting the SysML Activity Diagram from the SBC Interaction Transition Graph

In model-based systems engineering, the activity diagram defines the action in the activity along with the flow of input/output and control between them. The notion of an activity diagram is defined as follows.

**DEFINITION** (ACTIVITY DIAGRAM) A SysML activity diagram $AD = (\Psi, \Lambda, \Theta, \Gamma, ADR)$ consists of

- *a finite set $\Psi$ of states,*

- *a finite set $\Lambda$ of channel names,*

- *a finite set $\Theta$ of parameter lists,*

- *a finite set $\Gamma$ of blocks,*

- *a relation $ADR \subseteq \Psi_1$ X $\Lambda$ X $\Theta$ X $\Gamma$ X $\Psi_2$,*
  where $(s_j, ch, p, b, s_k) \in ADR$ *is denoted by* $s_j \xrightarrow{ch, p, b} s_k$ .

The SysML activity diagram of the system $AD_{system}$ is defined as $\big\|_{i=1, m} AD_i$ or $AD_1 \big\| AD_2 \big\| \dots \big\| AD_m$. Each activity diagram $AD_i$ is represented by the relation $ADR_i \subseteq \Psi_1$ X $\Lambda$ X $\Theta$ X $\Gamma$ X $\Psi_2$, where $(s_{ij}, ch, p, b, s_{ik}) \in ADR_i$ *is denoted by* $s_{ij} \xrightarrow{ch, p, b} s_{ik}$ . The activity diagram of the system $AD_{system}$ is represented by the relation $ADR_{system}$ which is defined as $\big\|_{i=1, m} ADR_i$ or $ADR_1 \big\| ADR_2 \big\| \dots \big\| ADR_m$, as shown in Table VI.

TABLE VI
Relation $ADR_{system}$

| $\Psi_1$ | $\Lambda$ | $\Theta$ | $\Gamma$ | $\Psi_2$ | | $\Psi_1$ | $\Lambda$ | $\Theta$ | $\Gamma$ | $\Psi_2$ |
|---|---|---|---|---|---|---|---|---|---|---|
| $s_{11}$ | $ch_{11}$ | $p_{11}$ | $b_{11}$ | $s_{12}$ | | $s_{21}$ | $ch_{21}$ | $p_{21}$ | $b_{21}$ | $s_{22}$ |
| $s_{12}$ | $ch_{12}$ | $p_{12}$ | $b_{12}$ | $s_{13}$ | $\big\|$ | $s_{22}$ | $ch_{22}$ | $p_{22}$ | $b_{22}$ | $s_{23}$ |
| • | • | • | • | • | | • | • | • | • | • |
| $s_{1n}$ | $ch_{1n}$ | $p_{1n}$ | $b_{1n}$ | $s_{11}$ | | $s_{2n}$ | $ch_{2n}$ | $p_{2n}$ | $b_{2n}$ | $s_{21}$ |

$\big\| \bullet \bullet \bullet \bullet \bullet \bullet \big\|$

| $\Psi_1$ | $\Lambda$ | $\Theta$ | $\Gamma$ | $\Psi_2$ |
|---|---|---|---|---|
| $s_{m1}$ | $ch_{m1}$ | $p_{m1}$ | $b_{m1}$ | $s_{m2}$ |
| $s_{m2}$ | $ch_{m2}$ | $p_{m2}$ | $b_{m2}$ | $s_{m3}$ |
| • | • | • | • | • |
| $s_{mn}$ | $ch_{mn}$ | $p_{mn}$ | $b_{mn}$ | $s_{m1}$ |

The algorithm used to project the AD relation $ADR_{system} \subseteq \Psi_1$ X $\Lambda$ X $\Theta$ X $\Gamma$ X $\Psi_2$ from the ITG relation $ITGR_{system} \subseteq \Psi_1$ X $\Xi$ X $\Lambda$ X $\Theta$ X $\Gamma$ X $\Psi_2$ is as follows.

**ALGORITHM 3** (Projecting $ADR_{system}$ from $ITGR_{system}$)

**For** i = 1, m **Loop**

SELECT $\Psi_1, \Lambda, \Theta, \Gamma, \Psi_2$ INTO $ADR_i$ FROM $ITGR_i$ ;

**End Loop;**

ORTHOGONALLY COMPOSE ALL $ADR_i$

(i.e., $\big\|_{i=1, m} ADR_i$ ) TO GET $ADR_{system}$

**END ALGORITHM**

Once we have the AD relation $ADR_{system}$, it is easy to get the SyML activity diagram of the system.

## IV. CASE: VENDING MACHINE

### A. Vending Machine System

A vending machine is an automated machine that sells products to customers. A customer inserts enough money for a product and selects a specific product; then, the vending machine dispenses the selected product and dispenses change if necessary. A vending machine is a rather complex system with several features. Customers interact with the coin receptacle and buttons of the vending machine and the vendor needs to refill the vending products and change. To create a system modeling of a vending machine, we begin by describing the requirements of the vending machine.

The vending machine accepts coins or credit from the customer in payment for their purchase. The vending machine returns the customer's payment if he or she decides not to make a product selection. The vending machine accepts the customer's product selection. Once the customer makes a selection, the vending machine dispenses the product to the customer. To ensure that most vending products are available for customers to purchase, the vendor regularly checks the product store to see if any product needs to be refilled. The vendor refills the product store as needed. The vendor regularly checks and refills the coin store as needed to ensure that there are deposited coins available for change.

### B. SBC Interaction Transition Graph of the Vending Machine

In C-M-SBC-PA, the semantics of the vending machine (VM) is represented by the SBC interaction transition graph $ITG_{VM}$ (defined as "$ITG_1 \big\| ITG_2 \big\| ITG_3 \big\| ITG_4 \big\| ITG_5$") as shown in Figure 4.

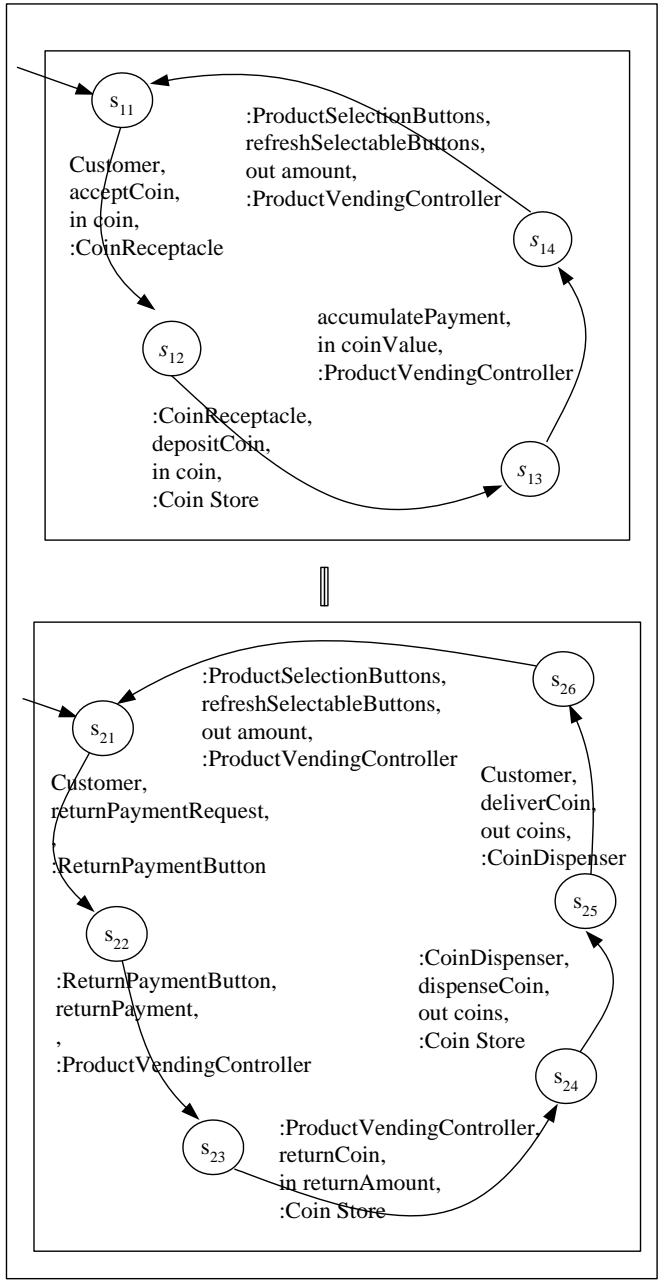

Figure 4. Transition Graph $ITG_{\text{VM}}$ (I)

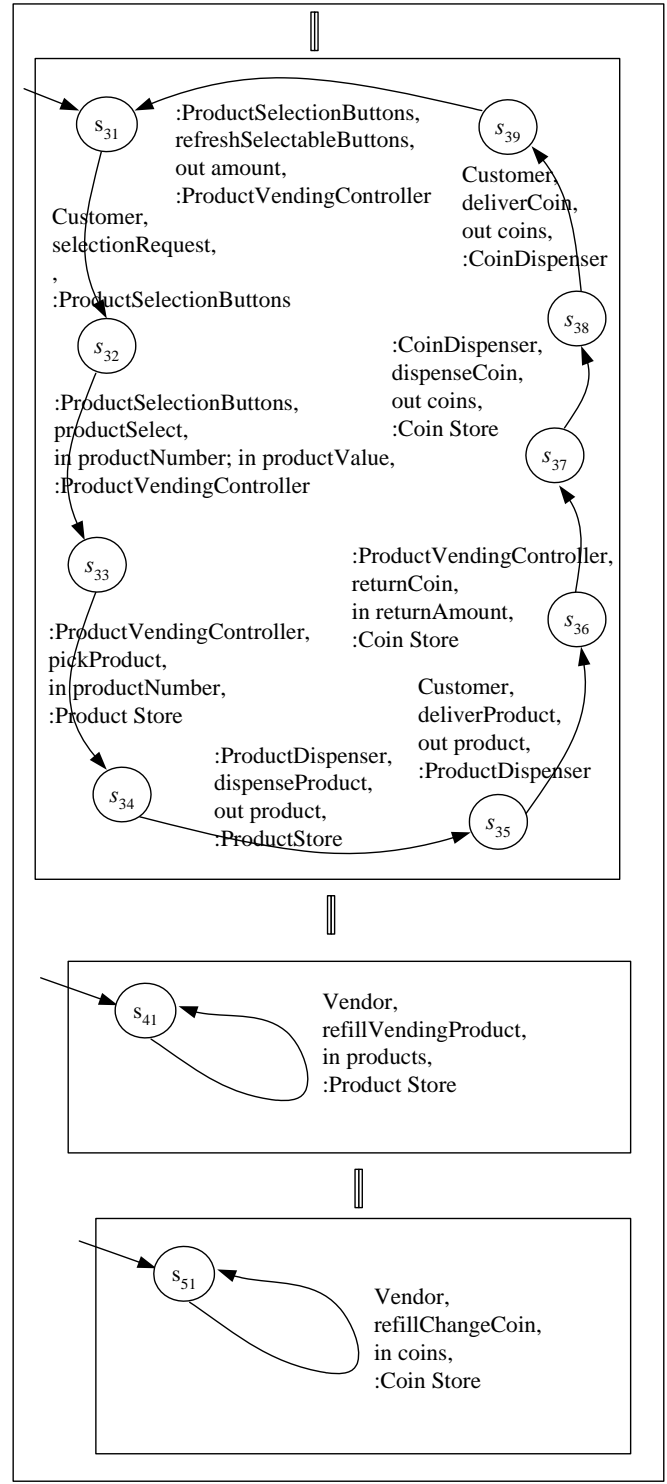

Figure 4. Transition Graph $ITG_{\text{VM}}$ (II)

We use the transition relation $ITGR_{\text{VM}} \subseteq \varPsi_1 \times \varXi \times \varLambda \times \varTheta \times \varGamma \times \varPsi_2$ (defined as "$ITGR_1 \| ITGR_2 \| ITGR_3 \| ITGR_4 \| ITGR_5$") to represent the SBC interaction transition graph of the vending machine, as shown in Table VII.

TABLE VII
Relation $ITGR_{\text{VM}}$ (I)

| $\Psi_1$ | $\Xi$ | $\Lambda$ | $\Theta$ | $\Gamma$ | $\Psi_2$ |
|---|---|---|---|---|---|
| $s_{11}$ | Customer | acceptCoin | in coin | :Coin Receptacle | $s_{12}$ |
| $s_{12}$ | :Coin Receptacle | depositCoin | in coin | :Coin Store | $s_{13}$ |
| $s_{13}$ | :Coin Receptacle | accumulate Payment | in coinValue | :Product Vending Controller | $s_{14}$ |
| $s_{14}$ | :Product Selection Buttons | refresh Selectable Buttons | out amount | :Product Vending Controller | $s_{11}$ |

| $\Psi_1$ | $\Xi$ | $\Lambda$ | $\Theta$ | $\Gamma$ | $\Psi_2$ |
|---|---|---|---|---|---|
| $s_{21}$ | Customer | return Payment Request | | :Return Payment Button | $s_{22}$ |
| $s_{22}$ | :Return Payment Button | return Payment | | :Product Vending Controller | $s_{23}$ |
| $s_{23}$ | :Product Vending Controller | returnCoin | in returnAmount | :Coin Store | $s_{24}$ |
| $s_{24}$ | :Coin Dispenser | dispense Coin | out coins | :Coin Store | $s_{25}$ |
| $s_{25}$ | Customer | deliverCoin | out coins | :Coin Dispenser | $s_{26}$ |
| $s_{26}$ | :Product Selection Buttons | refresh Selectable Buttons | out amount | :Product Vending Controller | $s_{21}$ |

TABLE VII
Relation $ITGR_{\text{VM}}$ (II)

| $\Psi_1$ | $\Xi$ | $\Lambda$ | $\Theta$ | $\Gamma$ | $\Psi_2$ |
|---|---|---|---|---|---|
| $s_{31}$ | Customer | selection Request | | :Product Selection Buttons | $s_{32}$ |
| $s_{32}$ | :Product Selection Buttons | product Select | in productNumber; in productValue | :Product Vending Controller | $s_{33}$ |
| $s_{33}$ | :Product Vending Controller | pickProduct | in productNumber | :Product Store | $s_{34}$ |
| $s_{34}$ | :Product Dispenser | dispense Product | out product | :Product Store | $s_{35}$ |
| $s_{35}$ | Customer | deliver Product | out product | :Product Dispenser | $s_{36}$ |
| $s_{36}$ | :Product Vending Controller | returnCoin | in returnAmount | :Coin Store | $s_{37}$ |
| $s_{37}$ | :Coin Dispenser | dispense Coin | out coins | :Coin Store | $s_{38}$ |
| $s_{38}$ | Customer | deliverCoin | out coins | :Coin Dispenser | $s_{39}$ |
| $s_{39}$ | :Product Selection Buttons | refresh Selectable Buttons | out amount | :Product Vending Controller | $s_{31}$ |

| $\Psi_1$ | $\Xi$ | $\Lambda$ | $\Theta$ | $\Gamma$ | $\Psi_2$ |
|---|---|---|---|---|---|
| $s_{41}$ | Vendor | refill Vending Product | in products | :Product Store | $s_{41}$ |

| $\Psi_1$ | $\Xi$ | $\Lambda$ | $\Theta$ | $\Gamma$ | $\Psi_2$ |
|---|---|---|---|---|---|
| $s_{51}$ | Vendor | refill Change Coin | in coins | :Coin Store | $s_{51}$ |

*C. Projecting the SysML Internal Block Diagram of the Vending Machine*

We apply the algorithm of projecting the IBD relation (i.e., $IBDR_{\text{VM}}$) from the ITG relation (i.e., $ITGR_{\text{VM}}$) of the vending machine. After the projection, we get the relation $IBDR_{\text{VM}} \subseteq \Xi \times \Lambda \times \Theta \times \Gamma$ as shown in Table VIII.

## TABLE VIII
### Relation $IBDR_{VM}$

| $\Xi$ | $\Lambda$ | $\Theta$ | $\Gamma$ |
|---|---|---|---|
| Customer | acceptCoin | in coin | :Coin Receptacle |
| :Coin Receptacle | depositCoin | in coin | :Coin Store |
| :Coin Receptacle | accumulate Payment | in coinValue | :Product Vending Controller |
| :Product Selection Buttons | refresh Selectable Buttons | out amount | :Product Vending Controller |
| Customer | return Payment Request | | :Return Payment Button |
| :Return Payment Button | return Payment | | :Product Vending Controller |
| :Product Vending Controller | returnCoin | in returnAmount | :Coin Store |
| :Coin Dispenser | dispense Coin | out coins | :Coin Store |
| Customer | deliverCoin | out coins | :Coin Dispenser |
| Customer | selection Request | | :Product Selection Buttons |
| :Product Selection Buttons | product Select | in productNumber; in productValue | :Product Vending Controller |
| :Product Vending Controller | pickProduct | in productNumber | :Product Store |
| :Product Dispenser | dispense Product | out product | :Product Store |
| Customer | deliver Product | out product | :Product Dispenser |
| Vendor | refill Vending Product | in products | :Product Store |
| Vendor | refill Change Coin | in coins | :Coin Store |

From the projected IBD relation $IBDR_{VM}$, we draw the corresponding SysML internal block diagram of the vending machine, as shown in Figure 5.

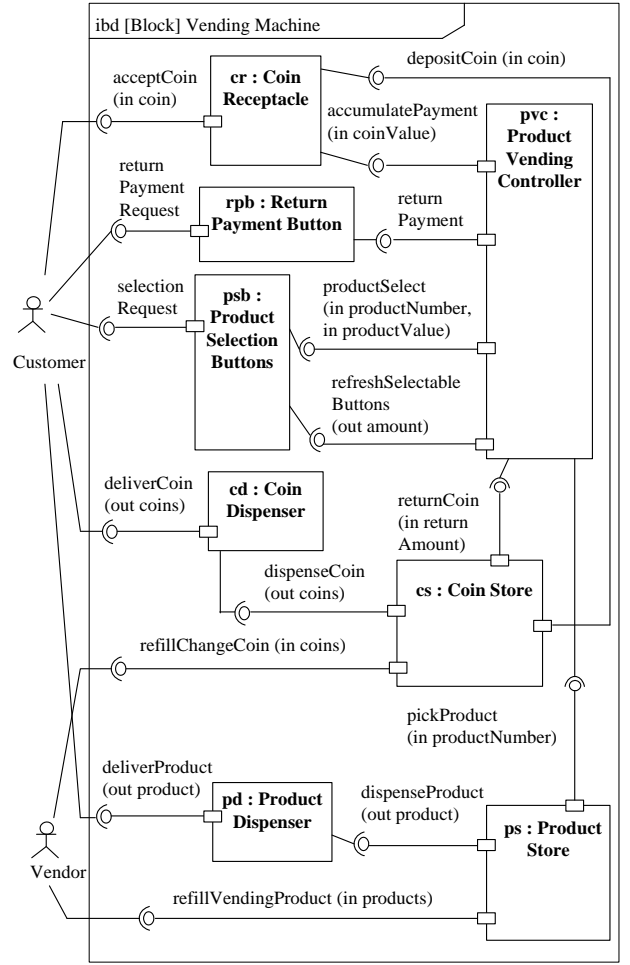

Figure 5. Projected IBD View of the Vending Machine

### D. Projecting the SysML State Machine Diagram of the Vending Machine

We apply the algorithm of projecting the SMD relation (i.e., $SMDR_{VM}$) from the ITG relation (i.e., $ITGR_{VM}$) of the vending machine. After the projection, we get the relation $SMDR_{VM} \subseteq \Psi_1 \times \Lambda \times \Psi_2$ as shown in Table IX.

TABLE IX
Relation $SMDR_{\text{VM}}$ (I)

| $\Psi_1$ | $\Lambda$ | $\Psi_2$ |
|---|---|---|
| $s_{11}$ | acceptCoin | $s_{12}$ |
| $s_{12}$ | depositCoin | $s_{13}$ |
| $s_{13}$ | accumulatePayment | $s_{14}$ |
| $s_{14}$ | refreshSelectableButtons | $s_{11}$ |

⫫

| $\Psi_1$ | $\Lambda$ | $\Psi_2$ |
|---|---|---|
| $s_{21}$ | returnPaymentRequest | $s_{22}$ |
| $s_{22}$ | returnPayment | $s_{23}$ |
| $s_{23}$ | returnCoin | $s_{24}$ |
| $s_{24}$ | dispenseCoin | $s_{25}$ |
| $s_{25}$ | deliverCoin | $s_{26}$ |
| $s_{26}$ | refreshSelectableButtons | $s_{21}$ |

⫫

TABLE IX
Relation $SMDR_{\text{VM}}$ (II)

| $\Psi_1$ | $\Lambda$ | $\Psi_2$ |
|---|---|---|
| $s_{31}$ | selectionRequest | $s_{32}$ |
| $s_{32}$ | productSelect | $s_{33}$ |
| $s_{33}$ | pickProduct | $s_{34}$ |
| $s_{34}$ | dispenseProduct | $s_{35}$ |
| $s_{35}$ | deliverProduct | $s_{36}$ |
| $s_{36}$ | returnCoin | $s_{37}$ |
| $s_{37}$ | dispenseCoin | $s_{38}$ |
| $s_{38}$ | deliverCoin | $s_{39}$ |
| $s_{39}$ | refreshSelectableButtons | $s_{31}$ |

⫫

| $\Psi_1$ | $\Lambda$ | $\Psi_2$ |
|---|---|---|
| $s_{41}$ | refillVendingProduct | $s_{41}$ |

⫫

| $\Psi_1$ | $\Lambda$ | $\Psi_2$ |
|---|---|---|
| $s_{51}$ | refillChangeCoin | $s_{51}$ |

From the projected SMD relation $SMDR_{\text{VM}}$, we draw the corresponding SysML state machine diagram of the vending machine, as shown in Figure 6.

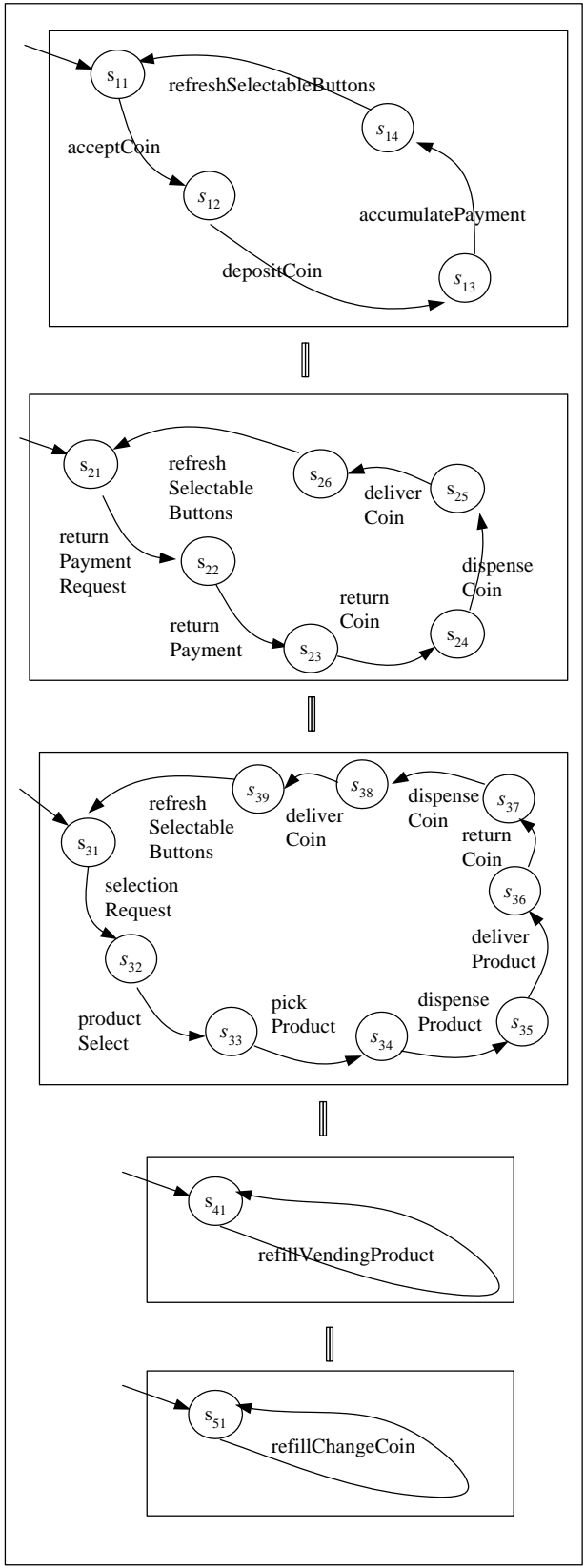

Figure 6.   Projected SMD View of the Vending Machine

*E.  Projecting the SysML Activity Diagram of the Vending Machine*

We apply the algorithm of projecting to the AD relation (i.e., $ADR_{\mathrm{VM}}$) from the ITG relation (i.e., $ITGR_{\mathrm{VM}}$) of the vending machine. After the projection, we get the relation $ADR_{\mathrm{VM}} \subseteq \Psi_1$ X $\Lambda$ X $\Theta$ X $\Gamma$ X $\Psi_2$ as shown in Table X.

TABLE X
Relation $ADR_{\mathrm{VM}}$ (I)

| $\Psi_1$ | $\Lambda$ | $\Theta$ | $\Gamma$ | $\Psi_2$ |
|---|---|---|---|---|
| $s_{11}$ | acceptCoin | in coin | :Coin Receptacle | $s_{12}$ |
| $s_{12}$ | depositCoin | in coin | :Coin Store | $s_{13}$ |
| $s_{13}$ | accumulate Payment | in coinValue | :Product Vending Controller | $s_{14}$ |
| $s_{14}$ | refresh Selectable Buttons | out amount | :Product Vending Controller | $s_{11}$ |

| $\Psi_1$ | $\Lambda$ | $\Theta$ | $\Gamma$ | $\Psi_2$ |
|---|---|---|---|---|
| $s_{21}$ | return Payment Request | | :Return Payment Button | $s_{22}$ |
| $s_{22}$ | return Payment | | :Product Vending Controller | $s_{23}$ |
| $s_{23}$ | returnCoin | in returnAmount | :Coin Store | $s_{24}$ |
| $s_{24}$ | dispense Coin | out coins | :Coin Store | $s_{25}$ |
| $s_{25}$ | deliverCoin | out coins | :Coin Dispenser | $s_{26}$ |
| $s_{26}$ | refresh Selectable Buttons | out amount | :Product Vending Controller | $s_{21}$ |

| $\Psi_1$ | $\Lambda$ | $\Theta$ | $\Gamma$ | $\Psi_2$ |
|---|---|---|---|---|
| $s_{31}$ | selection Request | | :Product Selection Buttons | $s_{32}$ |
| $s_{32}$ | product Select | in productNumber; in productValue | :Product Vending Controller | $s_{33}$ |
| $s_{33}$ | pickProduct | in productNumber | :Product Store | $s_{34}$ |
| $s_{34}$ | dispense Product | out product | :Product Store | $s_{35}$ |
| $s_{35}$ | deliver Product | out product | :Product Dispenser | $s_{36}$ |
| $s_{36}$ | returnCoin | in returnAmount | :Coin Store | $s_{37}$ |
| $s_{37}$ | dispense Coin | out coins | :Coin Store | $s_{38}$ |
| $s_{38}$ | deliverCoin | out coins | :Coin Dispenser | $s_{39}$ |
| $s_{39}$ | refresh Selectable Buttons | out amount | :Product Vending Controller | $s_{31}$ |

$\|$

| $\Psi_1$ | $\Lambda$ | $\Theta$ | $\Gamma$ | $\Psi_2$ |
|---|---|---|---|---|
| $s_{41}$ | refill Vending Product | in products | :Product Store | $s_{41}$ |

$\|$

| $\Psi_1$ | $\Lambda$ | $\Theta$ | $\Gamma$ | $\Psi_2$ |
|---|---|---|---|---|
| $s_{51}$ | refill Change Coin | in coins | :Coin Store | $s_{51}$ |

From the projected AD relation $ADR_{VM}$, we draw the corresponding SysML activity diagram of the vending machine, as shown in Figure 7.

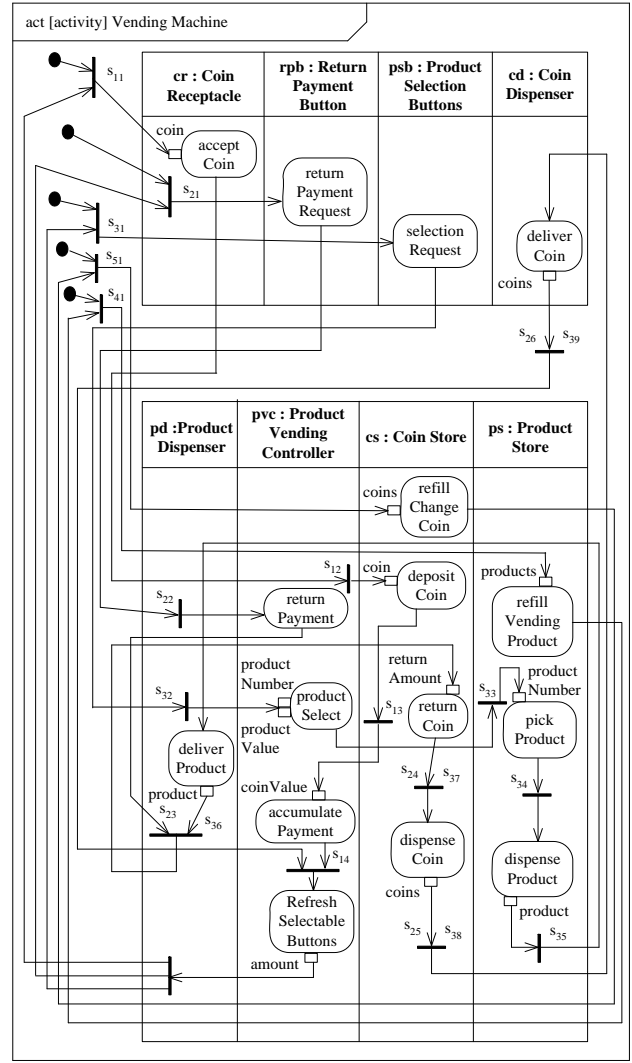

Figure 7. Projected AD View of the Vending Machine

## V. CONCLUSIONS

In this paper, Channel-Based Multi-Queue Structure-Behavior Coalescence Process Algebra (C-M-SBC-PA) is proposed as a metamodel for SysML in model-based systems engineering. One important use of the SysML metamodel is to provide an integrated semantic framework that every diagram in the user model can be projected as a view of the metamodel. Nowadays, most current SysML metamodels fail to project each diagram in the user model as a view of the metamodel.

In order to overcome the shortcomings of the current SysML metamodel approach, we need an integrated semantic framework that is able to unify the structural and behavioral constructs. Adopting C-M-SBC-PA as a metamodel for SysML, we use the SBC interaction transition graph as a single diagram to complete the overall semantic specification of the system. Through the SBC interaction transition graph and its corresponding SBC interaction transition graph relation, each diagram in the SysML user model can be projected as a view of the SBC interaction transition graph. Therefore, we conclude that the SBC interaction transition graph used by C-M-SBC-PA method as a metamodel for SysML is indeed a basis for unification of different views of the system in Model-Based Systems Engineering.

### ACKNOWLEDGEMENTS

The author is thankful to the anonymous reviewers for their useful remarks, which help illuminate the nuances and sparked new ideas.

**William S. Chao** was born in 1954 in Taiwan and received his Ph.D. degree in information science from the University of Alabama at Birmingham, USA, in 1988. William worked as a computer scientist at GE Research and Development Center, from 1988 till 1991 and has been teaching at National Sun Yat-Sen University, Taiwan since 1992. His research covers: systems architecture, hardware architecture, software architecture, and enterprise architecture. Dr. Chao is a member of the Association of Enterprise Architects Taiwan Chapter and also a member of the Chinese Association of Enterprise Architects.